\documentstyle[psfig]{elsart}

\newcommand{\ii}{\mathrm{i}}
\newcommand{\dsl}{\ii \FMSlash{\partial}}
\newcommand{\asl}{g{\lambda^a\over 2} \FMSlash{A}^a} 

\begin{document}

\begin{frontmatter}

\title{The chiral phase transition in a random matrix model with 
molecular correlations}

\author[Heidelberg]{Tilo Wettig\thanksref{TW}},
\author[Frankfurt]{A.\ Sch\"afer\thanksref{AS}}, and
\author[Heidelberg]{H.\ A.\ Weidenm\"uller\thanksref{HAW}}
\address[Heidelberg]{Max-Planck-Institut f\"ur Kernphysik, Postfach
  103980, D-69029 Heidelberg, Germany}
\address[Frankfurt]{Institut f\"ur Theoretische Physik, J.\ W.\ Goethe
  Universit\"at Frankfurt, Postfach 111932, D-60054 Frankfurt am Main,
  Germany}
\thanks[TW]{E-mail address: tilo@pluto.mpi-hd.mpg.de}
\thanks[AS]{E-mail address: schaefer@th.physik.uni-frankfurt.de}
\thanks[HAW]{E-mail address: haw@pluto.mpi-hd.mpg.de}

\date{1 November 1995}

\begin{abstract}
The chiral phase transition of QCD is analyzed in a model combining
random matrix elements of the Dirac operator with specially chosen
non-random ones. The special form of the latter is motivated by the
assumption that the fermionic quasi-zero modes associated with
instanton and anti-instanton configurations determine the chiral
properties of QCD. Our results show that the degree of correlation
between these modes plays the decisive role. To reduce the value of
the chiral condensate by more than a factor of 2 about 95 percent of
the instantons and anti-instantons must form so-called molecules.
This conclusion agrees with numerical results of the Stony Brook
group.
\end{abstract}

\end{frontmatter}

One of the challenging tasks of hadron physics consists in
understanding the properties of the chiral phase transition in QCD,
and its relation to the deconfinement phase transition.  This
fascinating theoretical problem requires the consistent combination of
several basic concepts of field theory: Dynamical and spontaneous
symmetry breaking, non-trivial topological properties, confinement,
and a hierarchy of correlation lengths.  A deeper understanding of the
chiral phase transition is also urgently required for the analysis of
heavy ion collisions at very high energies.  Such experiments will be
done at RHIC and at LHC in the near future.

Because of its importance, this problem has received considerable
theoretical attention especially during the last decade.  Some papers
which are particularly relevant for our investigation are listed in
Refs.~\cite{Di1}--\cite{Jack95}.  We use the notation of Shuryak,
Verbaarschot, and Jackson \cite{SV,Verb94,Jack95} whose papers
dominate the field. The derivation of the fundamental relations can be
found there.

The starting point for most discussions is the Banks-Casher formula
\cite{BC},
\begin{equation}
\langle \bar q q \rangle = -{\pi\over V_4} \rho(\mu)\Big|_{\mu=0} \;.
\end{equation} 
It links the value of the chiral condensate, $\langle \bar q q
\rangle$, to the density of the eigenvalues $\mu$ of the Dirac
equation,
\begin{equation} 
\left( \dsl+ \asl \right)\psi(x)=\mu \psi(x) \;, 
\label{Dirac}
\end{equation}
at $\mu=0$.  Here, $V_4$ is the four-volume, and $A^a$ the gluon field
operator.  The virtuality of typical fermion states is of order
$\Lambda_{\mathrm{QCD}}$ so that they do not contribute to $\rho(0)$.
This led to the assumption that $\rho(0)$ is dominated by very special
states, the zero modes.  Via the Atiyah-Singer index theorem,
\begin{equation}
{g^2\over 32\pi^2} \epsilon_{\mu\nu\rho\sigma} ~\int \d^4x ~
F^a_{\mu\nu}F^a_{\rho\sigma} = N_+-N_-\;,
\label{Atiyah}
\end{equation}
these modes are linked to topologically non-trivial gauge field
configurations (instantons and anti-instantons).  Here, $N_+$ ($N_-$)
is the number of zero modes with positive (negative) chirality.  The
integral on the left hand side contains the field tensor
$F^a_{\mu\nu}$ and yields the topological charge $Q$.  This charge is
$+1$ ($-1$) for an isolated instanton (anti-instanton).  To understand
the behavior of the chiral condensate (and, thus, of hadron masses) as
a function of, e.g., temperature $T$ it would then suffice to obtain
the statistical properties of the zero modes associated with
QCD-instantons. (For finite instanton separation, these are more
precisely refered to as quasi-zero modes.) Some time ago, it was
realized \cite{SV,Verb94} that at $T=0$, this statistical problem can
be analyzed with the help of random matrix models; recently, this
analysis was extended for the first time to finite temperature
\cite{Jack95}.

In this paper, we generalize these earlier contributions by
investigating the dependence of the condensate on two parameters which
characterize the tem\-pe\-ra\-tu\-re-dependence of the instanton
configuration.  One of these, $D$, is the strength of the (diagonal)
interaction between fermionic states of opposite chirality.  It is due
to the formation of instanton--anti-instanton pairs (``molecules'')
with an optimally attractive relative orientation in color space.
This parameter was also used in Ref.~\cite{Jack95}.  The second
parameter, $\alpha$, describes the fraction of instantons and
anti-instantons which form such molecules. The inclusion of this new
parameter is important because numerical simulations and
phenomenological considerations suggest that it plays a crucial role
in chiral symmetry breaking \cite{Ilge94}.  In some of our present
work, we actually go beyond this simple two-parameter model and
investigate distributions of the diagonal matrix elements.  In this
way, we derive some general statements about properties of the chiral
phase transition, and about the associated critical exponents.

We see from Eq.~(\ref{Atiyah}) that each instanton or anti-instanton
is associated with {\em one\/} quasi-zero mode.  In any given sector
of the theory, characterized by the value of the topological charge
$Q=N_+-N_-$, the difference between the number of instantons and of
anti-instantons is given by $Q$.  In this sector, the QCD partition
function for a theory with $N_f$ fermionic flavors is thus
approximated by
\begin{equation}
  \label{eq1}
  Z_{\mathrm{QCD}}(Q)=\lim_{N_{\pm}\rightarrow \infty} 
  \left\langle \prod_{f=1}^{N_f} \det
  \left(\dsl+\asl+\ii m_f \right)\right\rangle_A \;.
\end{equation}
Here, only quasi-zero modes are taken into account; the matrix in
(\ref{eq1}) accordingly has dimension $(2N+Q)$.  (We define $N=N_-$
and assume without loss of generality that $Q\geq 0$.)  The average in
(\ref{eq1}) extends over all gauge field configurations $A^a$ with
topological charge $Q$ and the usual measure.  The symbol $m_f$
denotes the mass term.

In a basis of chirality eigenstates, the operator $\dsl+\asl$ couples
fields of different chirality while the mass term in the determinant
conserves chirality.  The determinant can be rearranged so that the
first $(N+Q)$ states have positive chirality and are localized at
the $(N+Q)$ instanton positions, while the remaining $N$ states
are the quasi-zero states with negative chirality and are localized at
the anti-instanton positions.  This yields
\begin{equation}
  \det\left(\matrix{\ii m_f&\dsl+\asl\cr (\dsl+\asl)^{\dagger} &
    \ii m_f}\right) \;.
\end{equation} 
The chirality-changing matrix elements have the form
\begin{eqnarray}
  \lefteqn{\phi^*_{I_k} \left( \ii\!\!\stackrel{\rightarrow}
     {\FMSlash{\partial}}+\asl(I_k)+\asl(\bar I_l)+ \sum_{j\neq
     k,l}\asl(j)+\asl_{\mathrm{fluct}}\right)\phi_{\bar I_l}}
     \nonumber \\ 
  && = \phi^*_{I_k} \left( \asl(I_k)+ \sum_{j\neq k,l}\asl(j)
     +\asl_{\mathrm{fluct}}\right)\phi_{\bar I_l}\nonumber \\ 
  && = \phi^*_{I_k} \left(-\ii\!\!\stackrel{\leftarrow}
     {\FMSlash{\partial}} + \sum_{j\neq k,l}\asl(j) +
     \asl_{\mathrm{fluct}}\right)\phi_{\bar I_l} \;.
\label{eq1a}
\end{eqnarray}
Here, we have assumed that the $\phi_{I_k}$ ($\phi_{\bar I_l}$) are
exact zero modes of the Dirac operator when the gauge field $A^a$ in
Eq.~(\ref{Dirac}) is replaced by the field $A^a(I_k)$ [$A^a(\bar
I_l)$] due to a single instanton (anti-instanton).  It is assumed that
these instantons and anti-instantons differ and are well separated,
with uncorrelated color orientation.  The symbol
$A^a_{\mathrm{fluct}}$ denotes the fluctuating part of the gauge field
not linked to specific instanton configurations.  The matrix elements
in the last line of Eq.~(\ref{eq1a}) can be replaced by the elements
of a random matrix with suitable symmetry properties
\cite{SV,Verb94,Jack95}.  This leads to a random-matrix model with
partition function
\begin{eqnarray}
  Z_{\beta}(Q)=\int{\cal D}[W] &&\exp\left[-\frac{N\beta\Sigma^2}{2}
  \mathrm{tr}(WW^{\dagger})\right] \nonumber \\ &&\times
  \prod_{f=1}^{N_f}\det\left(\matrix{m_f&\ii W\!+\!\ii D\cr \ii
    W^{\dagger}\!+\!\ii D^{\dagger}&m_f}\right) \;.
  \label{eq2}
\end{eqnarray}
The matrix $W$ is a rectangular matrix with $(N+Q)$ rows and $N$
columns.  Depending on the underlying symmetry, we distinguish three
random matrix ensembles pertaining to the partition function
(\ref{eq2}), labeled by $\beta=1$, 2, and 4: the Gaussian orthogonal
ensemble (GOE, $\beta=1$), the Gaussian unitary ensemble (GUE,
$\beta=2$), and the Gaussian symplectic ensemble (GSE, $\beta=4$) with
real, complex, or quaternion real matrices $W$, respectively.  As
shown by Verbaarschot \cite{Verb94}, these three cases correspond to
the following gauge groups: SU(2) in the fundamental representation
for $\beta=1$, SU($N_c$) with $N_c\geq 3$ in the fundamental
representation for $\beta=2$, and non-Abelian gauge groups SU($N_c$)
for all $N_c$ with fermions in the adjoint representation for
$\beta=4$.  In this letter, we are concerned with the case $\beta=2$
only.  The integration measure in Eq.~(\ref{eq2}) is the Haar measure.
The parameter $\Sigma$ is a real number which is related to the value
of the condensate at $T=0$ and which we take to be independent of
temperature.  Lattice calculations \cite{DiG} suggest that $\Sigma$
can depend on temperature only weakly; such a dependence cannot affect
our results qualitatively.

The matrix $D$ is real and diagonal.  As mentioned above, $D$ is
included to describe the non-random components of the Dirac matrix
elements due to the formation of instanton--anti-instanton pairs with
an optimally attractive relative orientation in color space.  The
matrix $D$ is diagonal because only the two quasi-zero modes of
opposite chirality belonging to such a pair are coupled by a
non-random matrix element.  We expect the elements of $D$ to grow with
increasing temperature.  This point will be substantiated below.  The
$T$-dependence of $D$ generates the $T$-dependence of the chiral
condensate.  In the sequel, we study several distributions of the
matrix elements of $D$.  Analytical results are obtained only for the
simplest of these, for which a fraction $\alpha$ of the elements of
$D$ is non-zero, all non-zero elements being equal.  We believe that
the distributions we study should account for the most important
features of QCD related to the chiral phase transition, with one
possible proviso.  It would seem much more realistic to replace the
random matrix $W$ by a banded random matrix \cite{brm}.  This would
account for the fact that the interaction of widely separated zero
modes should be suppressed, and would introduce a correlation length
in the theory.  Work on this problem is under way.

We turn to the details of the calculation.  The determinant in
(\ref{eq2}) can be rewritten as an integral over anti-commuting
(Grassmann) variables \cite{Verb85},
\begin{eqnarray}
  Z_2(Q)=\int{\cal D}[W]{\cal D}[\psi]\exp\Bigg[\!\!&&-
  \sum_{f=1}^{N_f}\psi^{f*}\left(\matrix{m_f&\ii W\!+\!\ii D\cr
  \ii W^{\dagger}\!+\!\ii D^{\dagger}&m_f}\right)\psi^f \nonumber \\
  &&-N\Sigma^2\mathrm{tr}(WW^{\dagger})\Bigg] \;.
  \label{eq3}
\end{eqnarray}
The (anti-commuting) vector $\psi$ has $N_f\,(2N+Q)$ components
which we label as follows.  The upper index $f\!=\!1\ldots N_f$
denotes the flavor, and the lower index numbers the (anti-) instanton
zero modes.  The first $N+Q$ components of the lower index denote
the instanton zero modes labeled by $I_k$ with $k\!=\!1\ldots N+Q$,
and the remaining $N$ components the anti-instanton zero modes labeled
by $\bar I_l$ with $l\!=\!1\ldots N$.  Performing the integration over
$W$ we obtain
\begin{eqnarray}
  \label{eq5}
  Z_2(Q)=\int{\cal D}[\psi]\exp\Biggl[ \!\! && 
  \frac{1}{N\Sigma^2} \psi_{{\bar I}_l}^{f*}\psi_{{\bar I}_l}^{g}
                      \psi_{I_k}^{g*}\psi_{I_k}^{f}
  -m_f(\psi_{I_k}^{f*}\psi_{I_k}^f+
       \psi_{{\bar I}_l}^{f*}\psi_{{\bar I}_l}^f) \nonumber \\
  &&-\ii D_{l}(\psi_{I_l}^{f*}\psi_{{\bar I}_l}^f+
              \psi_{{\bar I}_l}^{f*}\psi_{I_l}^f) 
  \Biggr] \;.
\end{eqnarray}
Summation over repeated indices is assumed, and we have set
$D_{ll}=D_l$.  The four-fermion terms can be removed by means of a
Hubbard-Stratonovitch transformation at the expense of introducing
$2N_f^2$ additional integration variables.  These can be taken to be
the elements of a complex $N_f\times N_f$ matrix $S$.  The result can
be rearranged as
\begin{eqnarray}
  \label{eq7}
  Z_2(Q)=\int{\cal D}[S]{\cal D}[\psi]\exp\Biggl[ \!\! && 
  -N\Sigma^2\mathrm{tr}(SS^{\dagger})
  -\sum_{k=N+1}^{N+Q} \psi_{I_k}^* (S+M) \psi_{I_k} \nonumber \\
  &&-\sum_{k=1}^{N}\left(\matrix{\psi_{I_k}^* \cr \psi_{\bar I_k}^*}\right)
  \left(\matrix{S+M & \ii D_k \cr \ii D_k & S^{\dagger}+M}\right)
  \left(\matrix{\psi_{I_k} \cr \psi_{\bar I_k}}\right)
  \Biggr] \;,
\end{eqnarray}
where the $\psi$-vectors are now in flavor space, the $D_k$ are
multiplied by the $N_f\times N_f$ unit matrix, and the matrix $M$ is a
diagonal matrix with entries $m_f$.  The integration over the
Grassmann variables can now be performed to yield
\begin{eqnarray}
  \label{eq8}
  Z_2(Q)=\int{\cal D}[S]&&\exp\left[-N\Sigma^2\mathrm{tr}(SS^{\dagger})
  \right]{\det}^{Q}(S+M) \nonumber \\
  &&\times\prod_{k=1}^{N}\det\left(\matrix{S+M & \ii D_k \cr 
  \ii D_k & S^{\dagger}+M}\right) \;.
\end{eqnarray}

The $\langle\bar qq\rangle$-condensate is given by
\begin{equation}
  \label{eq8a}
  \langle\bar qq\rangle=\frac{1}{V_4N_f} 
  \left.\frac{\partial\log Z}{\partial M}\right|_{M=0} 
\end{equation}
which, using (\ref{eq8}), can be expressed as
\begin{eqnarray}
  \label{eq8b}
  \langle\bar qq\rangle=\int{\cal D}[S]&&\exp\left[-N\Sigma^2
  \mathrm{tr}(SS^{\dagger}) \right]{\det}^{Q}(S)
  \prod_{k=1}^{N}\det\left(\matrix{S & \ii D_k \cr \ii D_k &
    S^{\dagger}}\right) \nonumber \\ &&\times\frac{1}{V_4N_f}\left\{
  Q\mathrm{tr}(S^{-1})+ \sum_{k=1}^N\mathrm{tr}\left( \matrix{S& \ii
    D_k \cr \ii D_k & S^{\dagger}}\right)^{-1}\right\} \;.
\end{eqnarray}

In the limit $N\rightarrow\infty$, the integral in (\ref{eq8}) can be
evaluated in saddle-point approximation.  Neglecting terms of order
$Q$, we obtain
\begin{equation}
  \label{eq9}
  N\Sigma^2 S = \sum_{k=1}^N (S+M)\left[(S+M)(S^{\dagger}+M)+D_k^2
                \right]^{-1} \;.
\end{equation}
Any square complex matrix of dimension $N_f$ can be written as
$S=U\Lambda V^{-1}$, where $\Lambda$ is a diagonal matrix with real
and non-negative entries, $U$ is unitary, and
$V\!\in\!\mathrm{U}(N_f)/\bigotimes_{n=1}^{N_f}\mathrm{U}(1)$.  The
solution of (\ref{eq9}) in the limit of zero masses is, thus, obtained
by solving
\begin{equation}
  \label{eq10}
  \Sigma^2 = \frac{1}{N}\sum_{k=1}^N \frac{1}{\lambda^2+D_k^2}
\end{equation}
for $\lambda$.  The $\langle\bar qq\rangle$-condensate then follows
from Eq.~(\ref{eq8b}),
\begin{equation}
  \label{eq11}
  \langle\bar qq\rangle=\Sigma^2\lambda {2N+Q\over V_4} =
  \Sigma^2\lambda n_{\rm inst} \;,
\end{equation}
where $n_{\rm inst}$ is the instanton--anti-instanton density.

In a realistic physical situation, the entries $D_k$ are
distributed according to some (normalized) distribution $P(D_k)$.  In
the limit $N\rightarrow\infty$, we can map the (discrete) interval
$k\!=\!1\ldots N$ onto the (continuous) interval $x\!=\!0\ldots 1$.
The desired value of $\lambda$ is then obtained by solving the
equation
\begin{equation}
  \label{eq12}
  \Sigma^2 = \int_0^1 \frac{\d x}{\lambda^2+D^2(x)} 
\end{equation}
for $\lambda$.   Here, $D(x)$ is defined by
\begin{equation}
  \label{eq13}
  \int_0^{D(x)} P(y) \d y = x \;. 
\end{equation}

Although the above equations cannot in general be solved analytically
one can still make interesting statements about the connection between
the existence of a condensate and the analytical form of $P(y)$ for
small $y$.  A condensate exists if Eq.~(\ref{eq12}) has a positive
solution for $\lambda$.  This is the case if $\int_0^1 \d
x/D^2(x)>\Sigma^2$.  Since $D(x)$ increases monotonically by
definition this condition is automatically fulfilled if $D(x)\sim x^a$
for small $x$ with $a\geq 1/2$.  Using Eq.~(\ref{eq13}), this is
equivalent to $P(y)\sim y^b$ for small $y$ with $b\leq 1$.  The
distribution $P$ is a function of temperature.  Hence, necessary
conditions for the vanishing of the condensate above some critical
temperature $T_{\mathrm{c}}$ are
\begin{eqnarray}
  \label{eq14}
  &&P(y) \sim y^b \ \ \mathrm{for} \ \ y\rightarrow 0 \ \ 
  \mathrm{with} \ \ b>1 \\ 
  \label{eq15}
  &&\int_0^1 \frac{\d x}{D^2(x)} \leq \Sigma^2 \;.
\end{eqnarray}

While these conditions set non-trivial constraints on the general
distribution of the diagonal elements, it is more convenient for our
purposes to consider a number of interesting special cases where
analytical results can be obtained. The simplest possible case is that
where all elements of $D$ are equal, corresponding to
$P(y)=\delta(y-d)$ and $D(x)=d$.  The solution for $\lambda$ can be
obtained immediately to yield
\begin{equation}
  \label{eq16}
  \langle\bar qq\rangle=\Sigma n_{\rm inst}\sqrt{1-(\Sigma d)^2} \;.
\end{equation}
This version of the model with $d=\pi T$ was considered by Jackson and
Verbaarschot \cite{Jack95}.

As a more realistic case we assume that a fraction $\alpha\in [0,1]$
of the elements of $D$ are equal to a common value $d$ and the
remaining elements are zero.  This corresponds to
$P(y)=\alpha\delta(y-d)+(1-\alpha) \delta(y)$, $D(x)=0$ for
$0<x<1-\alpha$, and $D(x)=d$ for $1-\alpha<x<1$.  Again, we can solve
for $\lambda$ to obtain
\begin{equation}
  \label{eq17}
  \langle\bar qq\rangle=\frac{\Sigma n_{\rm inst}}
{\sqrt{2}}\sqrt{1-(\Sigma d)^2
    +\sqrt{[1-(\Sigma d)^2]^2+4(\Sigma d)^2(1-\alpha)}} \;.
\end{equation}
This result is shown graphically in Fig.~1.  Its most important
property is that $\langle \bar qq \rangle$ can approach zero only if
$\alpha$ is very close to one. This observation is in striking
agreement with the results of a specific model by Ilgenfritz and
Shuryak \cite{Ilge94} and with the general observation that the chiral
condensate changes abruptly at the chiral phase transition \cite{Ka}
while the instanton density changes only very smoothly.  Note also
that the relatively sharp drop of the chiral condensate close to the
critical temperature is in qualitative agreement with results obtained
by the Bern Group in the framework of chiral perturbation theory
\cite{Gass87}.  We conclude from our results that this is a generic
property which follows from the symmetries of the interaction and the
fundamental properties of instanton molecules.  Thus, any specific
model should reproduce these features.
\begin{figure}[htbp]
\centerline{\psfig{figure=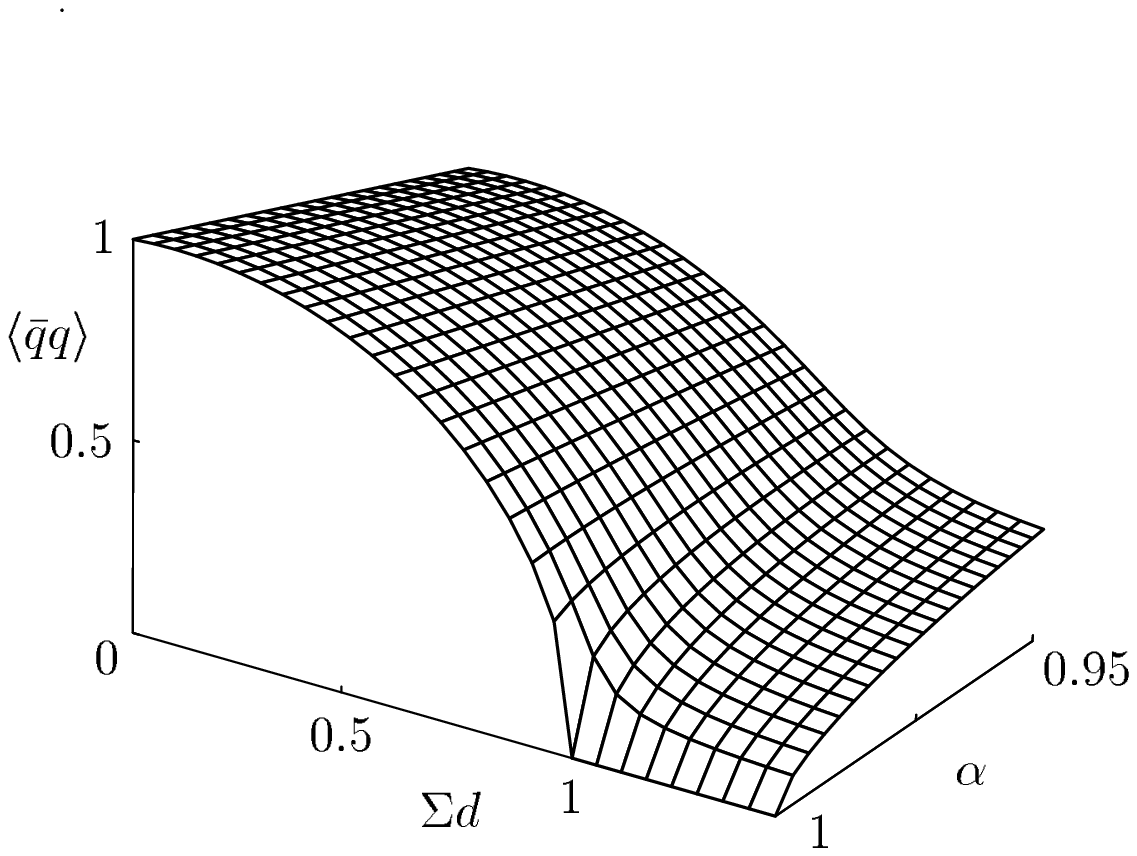,width=120mm}}
\caption{Plot of the $\langle \bar q q\rangle$ condensate in units of
  its value at $T=0$ as a function of $d$ and $\alpha$ defined in the
  text.  Both $d$ and $\alpha$ are functions of temperature so that
  the dependence of $\langle \bar q q\rangle$ on $T$ corresponds to a
  specific path on this surface.}
\end{figure}

At this point, we are in a position to discuss specific numbers
corresponding to the real physical situation.  The QCD vacuum state is
characterized by a number of quantities, one of them being the gluon
condensate $\langle {\alpha\over\pi} G^a_{\mu\nu}
G^a_{\mu\nu}\rangle$, which in turn can be related to the
instanton--anti-instanton density (in Euclidean space-time).
Phenomenologically, $n_{\rm inst}\approx$ 1~fm$^{-4}$ at $T=0$.  It is
expected \cite{Ilge94} that this value changes only little with
temperature up to $T_{\mathrm{c}}$.  The typical instanton radius
$\rho$ is equal to the color correlation length which is determined by
many phenomenological considerations (e.g., from pomeron phenomenology
or from the gluon form factor \cite{AS}) to be about 0.3~fm at $T=0$.
The same number is determined by fitting a host of hadronic properties
in the instanton vacuum model of Shuryak and co-workers.  These
numbers are further corroborated by specialized lattice studies using
cooling techniques to single out the quasi-classical field
configurations \cite{Chu,Chu95}.  Furthermore, the separation of
instanton and anti-instanton in a correlated pair is $R\approx$ 1~fm
\cite{Chu}, and the empirical value of the quark condensate is
$\langle \bar qq \rangle \approx -$(225 MeV)$^3$ \cite{Rein85}.

The diagonal matrix elements $d$ have been calculated by Diakonov and
Petrov \cite{Di1a}, Eq.~(22). For an optimally attractive orientation
of instanton and anti-instanton this implies
\begin{equation}
  |d|={4\rho^2\over R^3}\approx 0.4~{\rm fm}^{-1}
\end{equation}
at $T=0$.  On the other hand, we have for the ground state with no
noticeable instanton-anti-instanton correlation ($\alpha\rightarrow
0$)
\begin{equation}
  \Sigma = \left. {\langle \bar q q \rangle \over n_{\rm inst}}
  \right|_{T=0} \approx -1.5~{\rm fm} \;.
\end{equation}
These numbers imply $\sqrt{1-(\Sigma d)^2}\approx 0.8$ at $T=0$. Thus,
it is not sufficient that $\alpha$ approaches one, but in addition $d$
has to increase by about fifty percent as $T$ increases from zero to
$T_{\rm c}$.  This can be achieved by an increase of $\rho$ and/or a
decrease of $R$.  Lattice studies indicate that $\rho$ is
approximately independent of $T$ \cite{Chu95}.  Recent calculations
by T.\ Sch\"afer and E.\ Shuryak in the framework of the instanton
liquid model show that $R$ decreases with $T$ \cite{Scha95a}.  Thus,
an overall inrease of $d$ with temperature is plausible.

We can compute various critical exponents from the above results.  Our
order parameter is $\langle \bar q q\rangle$, and the control
parameter is the temperature $T$.  Let us concentrate on
Eq.~(\ref{eq17}).  The two parameters $d$ and $\alpha$ depend on $T$
in a way which is beyond the scope of our model.  However, some
qualitative statements can be made on physical grounds.  At $T=0$, $d$
has some finite value whereas $\alpha$ is zero.  Both parameters will
grow with temperature, but $\alpha$ is bounded by one.  It is
reasonable to assume that $d$ is an analytic function of $T$, but no
such assumption can be made for the behavior of $\alpha(T)$ close to
$\alpha=1$.  There is even the possibility that $\alpha$ changes
discontinuously to one resulting in a first order transition.  We note
that our model is rich enough to allow for that possibility but will
assume in the following discussion that $\alpha$ is a continuous
though not necessarily analytic function of $T$.

It is clear from Eq.~(\ref{eq17}) and Fig.~1 that necessary conditions
for a vanishing condensate are $\Sigma d \geq 1$ and $\alpha=1$.  Let
us denote the temperatures at which these two conditions are first
satisfied by $T_d$ and $T_{\alpha}$, respectively.  The larger of
these two temperatures is the critical temperature $T_\mathrm{c}$, but
we do not know a priori which one it will be.  In principle,
there are three possibilities: (i) $T_\mathrm{c}=T_d>T_{\alpha}$, (ii)
$T_\mathrm{c}=T_d=T_{\alpha}$, and (iii)
$T_\mathrm{c}=T_{\alpha}>T_d$.  We are interested in the critical
exponent $\beta$ defined by
\begin{equation}
  \label{eq18}
  \langle \bar q q \rangle \sim |t|^{\beta} \ \ \mathrm{with}
  \ \ t=\frac{T-T_\mathrm{c}}{T_\mathrm{c}} \;.
\end{equation}
(This $\beta$ is, of course, different from the $\beta$ labeling the
random matrix ensembles.)  For cases (ii) and (iii), $\beta$ depends
on the specific form of $\alpha(T)$ close to $T_{\alpha}$.  We
parameterize this dependence by $1-\alpha(t)\sim |t|^x$ for
$t\rightarrow0^-$.  We then obtain for case (i) $\beta=\half$, for
case (ii) $\beta=\half$ for $x\geq 2$ and $\beta=\frac{x}{4}$ for
$x<2$, and for case (iii) $\beta=\frac{x}{2}$.

The critical exponents $\gamma$ and $\delta$ are defined by
\begin{eqnarray}
  &&\chi \sim |t|^{-\gamma} \ \ \mathrm{with} \ \ \chi \sim \left.
  \frac{\partial \langle \bar q q \rangle}{\partial m}\right|_{m=0} 
  \label{eq19} \\ 
  &&\langle \bar q q \rangle \sim m^{1/\delta} \ \ \mathrm{at} \ \ 
  T=T_\mathrm{c} \;. \label{eq20}
\end{eqnarray}
In this case, the mass $m$ plays the role of the conjugate field
breaking chiral symmetry explicitly (the analog of the external
magnetic field in the case of a ferromagnet).  Both $\gamma$ and
$\delta$ can be computed in a straightforward manner by going back to
Eq.~(\ref{eq9}).  The results, however, again depend on the
relationship between $T_d$ and $T_{\alpha}$.  We find that in case (i)
$\gamma=1$ both above and below the transition and $\delta=3$ in
agreement with Ref.~\cite{Jack95}.  In case (iii) $\gamma=0$ both
above and below the transition and $\delta=1$.  Case (ii) is more
complicated.  We obtain that above the transition $\gamma=1$ whereas
below the transition $\gamma=1$ for $x\geq 2$ and $\gamma=\frac{x}{2}$
for $x<2$.  Furthermore, $\delta=3$ here.  In all three cases, the
Widom scaling law, $\gamma=\beta(\delta-1)$, is obeyed.

Case (i) and case (ii) for $x\geq 2$ yield the standard mean-field
exponents while case (ii) for $x<2$ and case (iii) give rise to
critical exponents which do not seem to be in a familiar universality
class.  This finding deserves further discussion.

When considering a second order phase transition, one usually assumes
that the ``input'' parameters of the system (in our case, $d$ and
$\alpha$) depend on the control parameter in a smooth and analytic
way.  The non-analytic features of the ``output'' parameters (e.g.,
the order parameter) at the critical temperature are a consequence of
the phase transition.  It is not in the spirit of such considerations
to introduce non-analyticities at the outset, i.e., in the input
parameters.  In our specific case, we are not able to specify how
$\alpha$ approaches one as $T\rightarrow T_{\alpha}^-$.  However,
$\alpha$ is special in the sense that it is strictly equal to one for
$T>T_{\alpha}$.  Thus, requiring $\alpha(T)$ to be continuous and
differentiable at $T_{\alpha}$ yields $1-\alpha(T)\sim
(T_{\alpha}-T)^2$ for $T\rightarrow T_{\alpha}^-$ and, hence, $x=2$ in
the above results for $\beta$ and $\gamma$.  Therefore, if it should
turn out that either $T_d>T_{\alpha}$ or $T_d=T_{\alpha}$ with
$\alpha(T)$ continuous and differentiable at $T_{\alpha}$, we
reproduce the mean field exponents of a second order phase transition.
This would be consistent with the recent discussion in \cite{Koci95}.
If $T_d<T_{\alpha}$ or $T_d=T_{\alpha}$ with $\alpha(T)$ continuous
but not differentiable at $T_{\alpha}$, we obtain a second order phase
transition with non-standard critical exponents.  Finally, if $\alpha$
is discontinuous at $T_{\alpha}\geq T_d$, we obtain a first order
transition.

In summary, we have extended the chiral random-matrix model of
Refs.~\cite{SV,Verb94,Jack95} in such a way that the model allows
either for a first-order or for a second-order chiral phase
transition.  For the second-order transition, we have calculated the
critical exponents, and we found a number of possible solutions, all
consistent with universal scaling laws.  A decision between them is
beyond the present scope of our model.  We believe, however, that
these options define the generic possibilities of any physically
realistic model of the chiral phase transition.

We would like to thank A.\ D.\ Jackson and J.\ J.\ M.\ Verbaarschot
for communicating their results prior to publication and useful
discussions.  TW also acknowledges helpful discussions with T.\ Guhr,
T.\ Sch\"afer, and E.\ V.\ Shuryak.  AS thanks D.\ Diakonov for very
valuable discussions and the MPI f\"ur Kernphysik, Heidelberg, as well
as the ECT$^*$, Trento, for its support.


\begin{thebibliography}{99}

\bibitem{Di1} D.\ I.\ Diakonov and V.\ Yu.\ Petrov, 
  Phys.\ Lett.\ B 147 (1984) 351;
  Nucl.\ Phys.\ B 245 (1984) 259;
  Sov.\ Phys.\ JETP 62 (1985) 204, 431.
\bibitem{Di1a} D.\ I.\ Diakonov and V.\ Yu.\ Petrov, 
  Nucl.\ Phys.\ B 272 (1986) 457.
\bibitem{Di2}  D.\ I.\ Diakonov and A.\ D.\ Mirlin,
  Phys.\ Lett.\ B 203 (1988) 2991.
\bibitem{Si} Yu.\ A.\ Simonov,
  Sov.\ J.\ Nucl.\ Phys.\ 53 (1991) 1099;
  Phys.\ Rev.\ D 43 (1991) 3534.
\bibitem{SV} E. Shuryak and J. Verbaarschot,
  Nucl.\ Phys.\ A 560 (1993) 306;
  Nucl.\ Phys.\ B 341 (1990) 1;
  Phys.\ Rev.\ Lett.\ 68 (1992) 2576.
\bibitem{Ver}  J. Verbaarschot, Nucl.\ Phys.\ B 427 (1994) 534.
\bibitem{Verb94} J.\ J.\ M.\ Verbaarschot, 
  Phys.\ Rev.\ Lett.\ 72 (1994) 2531.
\bibitem{Ilge94} E.-M.\ Ilgenfritz and E.\ V.\ Shuryak, 
  Phys.\ Lett.\ B 325 (1994) 263.
\bibitem{Scha95} T.\ Sch\"afer, E.\ V.\ Shuryak, and J.\ J.\ M.\
  Verbaarschot, Phys.\ Rev.\ D 51 (1995) 1267.
\bibitem{DiG} A.\ DiGiacomo, E.\ Meggiolaro, and H.\ Panagopoulos,
  Phys.\ Lett.\ B 277 (1992) 491.
\bibitem{Ka} F.\ Karsch, 
  XI. Conference on Ultra-Relativistic Nucleus-Nucleus-Collisions, 
  Monterey 1995, USA, preprint BI-TP 95/11, hep-lat/9503010 (1995).
\bibitem{Ba} C.\ F.\ Baillie et al., 
  Phys.\ Lett.\ B 197 (1987) 195.
\bibitem{Bun} G.\ Bunatian and J.\ Wambach, 
  Phys.\ Lett. B 336 (1994) 290.
\bibitem{Jack95} A.\ D.\ Jackson and J.\ J.\ M.\ Verbaarschot,
   preprint hep-ph/9509324 (1995).
\bibitem{BC} T.\ Banks and A.\ Casher, 
  Nucl.\ Phys.\ B 169 (1980) 103.
\bibitem{brm} Y.\ V.\ Fyodorov and A.\ D.\ Mirlin,
  Phys.\ Rev.\ Lett.\ 67, 2405 (1991). 
\bibitem{Verb85} J.\ J.\ M.\ Verbaarschot, H.\ A.\ Weidenm\"uller, and
  M.\ Zirnbauer, 
  Phys.\ Rep.\ 129 (1985) 367.
\bibitem{Gass87} J.\ Gasser and H.\ Leutwyler, 
  Phys.\ Lett.\ B 188, 477 (1987). 
\bibitem{AS} V.\ M.\ Braun, P.\ Gornicki, L.\ Mankiewicz, and A.\
  Sch\"afer, Phys.\ Lett.\ B 302 (1993) 291.
\bibitem{Chu} M.-C.\ Chu, J.\ M.\ Grandy, S.\ Huang, and J.\ W.\
  Negele, Phys.\ Rev.\ D 49 (1994) 6039; 
  D.\ I.\ Diakonov, ``Instanton vacuum and confinement in QCD'', 
  Third St.\ Petersburg Winter School in QCD, Feb.\,26--Mar.\,11, 1995.
\bibitem{Chu95} M.-C.\ Chu and S.\ Schramm, Phys.\ Rev.\ D 51 (1995)
  4580.
\bibitem{Rein85} L.\ J.\ Reinders, H.\ Rubinstein, and S.\ Yazaki, 
  Phys.\ Rep.\ 127, 1 (1985).
\bibitem{Scha95a} T.\ Sch\"afer and E.\ V.\ Shuryak, preprint
  hep-ph/9509337 (1995). 
\bibitem{Koci95} A.\ Koci\'c and J.\ Kogut, 
  Phys.\ Rev.\ Lett.\ 74 (1995) 3109.
\end{thebibliography}
\end{document}